# Revisiting spin state crossover in (MgFe)O by means of high resolution X-ray diffraction from a single crystal.

Konstantin Glazyrin[1,*], Saiana Khandarkhaeva[2], Leonid Dubrovinsky[2] and Michael Sprung[1]

[1]*PhotonSciences, Deutsches Elektronen Synchrotron (DESY), 22607, Hamburg, Germany*

[2]*Bayerisches Geoinstitut, University of Bayreuth, Bayreuth, 95440, Germany*

[*] Contact author email: konstantin.glazyrin@desy.de

**ABSTRACT**

(MgFe)O is a solid solution with ferrous iron undergoing the high to low spin state (HS-LS) crossover under high pressure. The exact state of the material in the region of the crossover is still a mystery, as domains with different spin states may coexist over a wide pressure range without changing the crystal structure neither from the symmetry nor from the atomic positions point of view. At the conditions of the crossover, (MgFe)O is a special type of microscopic disorder system. We explore the influences of (a) stress-strain relations in a diamond anvil cell, (b) time relaxation processes, and (c) the crossover itself on the characteristic features of a single crystal $(111)_{hkl}$ Bragg spot before, during and after the transformation. Using high resolution X-ray diffraction as a novel method for studies of unconventional processes at the conditions of suppressed diffusion, we detect and discuss subtle changes of the $(111)_{hkl}$ Bragg spot projections which we measure and analyze as a function of pressure. We report changes of the spot shape which can be correlated with the HS-LS relative abundance. In addition, we report the formation of structural defects as an intrinsic material response. These static defects are accumulated during transformation of the material from HS to LS.

# I. INTRODUCTION

The solid state solution MgO-FeO (or (Mg,Fe)O) would seem trivial if not playing role for the lower mantle of the Earth (and probably super-Earths) and exhibit a complex and at the same time fascinating behavior. The spin state crossover of ferrous iron can indeed play an important role in natural processes of planetary scale. On the one hand the crossover affects elastic properties of the materials. On the other hand, it may be involved in various diverse processes featuring multiple phases, e.g. involving iron partitioning in major constituents of geophysically and geochemically important mineral assemblages (Earth's lower mantle, ultralow velocity zones, terrestrial and extraterrestrial planets, and etc.) [1–4].

At the same time we should not forget that (MgFe)O is an oxide system with strong correlation between lattice and spins. We would like to follow the original publication of E. Dagotto [5] and reference multifaceted complexity of the transition metal oxides for systems with dominant states which are competing, but not spatially homogeneous, and focus more on description of (MgFe)O as a strongly correlated system. We will focus mostly on the (MgFe)O with low to moderate concentration of iron (e.g. $x_{Fe}$=0.19(1) ), which remains in a cubic structure at all pressures relevant for the spin state crossover.

The spin state crossover in (MgFe)O is complex thermodynamic quantum phenomenon [1,6,7]. There is a consensus in literature that during the crossover the high and low spin states of ferrous iron concur in the material iso-structurally. The state of coexistence, in turn, should imply a creation of a spatially inhomogeneous disordered system with competing domains of $Fe^{2+}$ in high (HS) and low (LS) states. Review of the literature reflects the significant progress unraveling different important aspects of this phenomenon [1], still most of the studies are focused on the properties of the bulk material. Thus, the phenomenon and the underlying system were never closely investigated on the sub-micron and nano- scale. After a careful consideration it becomes clear that the origins behind the changes of the bulk are hidden on atomic level as we attempt to demonstrate below.

Within the pressure range of the spin state crossover, (MgFe)O can be introduced as a 'frustrated system'. While the total abundance of HS and LS domains is dictated by thermodynamic pressure and temperature conditions, the exact distribution of HS-LS domains should be free to change for a given thermodynamic point. With or without a small temperature perturbation, the system may switch a specific $Fe_{LS}^{2+}$ to $Fe_{HS}^{2+}$, individually. The restraint on the spin state abundance balance would require another $Fe_{HS}^{2+}$ to switch into $Fe_{LS}^{2+}$, promoting 'breathing' of a system. The presence of competing, but spatially heterogeneous dominant states could be described within the term of Griffiths phases, so common for other transition elements oxides (e.g. [8,9] citing papers and the references within).

It is well documented that in addition to pressure, the system is very sensitive to the temperature. Temperature increase expands the pressure region of HS and LS coexistence [6,10]. Thus, even under conditions of a strongly suppressed diffusion of Mg and Fe at high pressure and ambient temperature [11], even small temperature fluctuations introduce an additional complication to the picture.

The description of (MgFe)O spin state crossover as a frustrated system can be reconceived into a picture of the system disorder. At the moment, we do not have any information about the equilibrium shapes, sizes and the distribution of HS and LS domains. They may be small, down to nanoscale size. They could be still large enough (i.e. 10-100 of nm) that they would not create a significant detectable contribution to the conventional high pressure powder [12] or single crystal diffraction signal from individual spin states (Supplementary of Ref. [13]). What is important, is that these domains have significant contact strain, e.g. separating configurations with different spin states. The HS and LS domains of (Mg,Fe)O should have the lower and the higher densities, respectively, and, thus, their presence and competition introduce a density fluctuation on atomic scale (i.e. with a simplest picture of a single $Fe_{LS}^{2+}$ surrounded by six $Fe_{HS}^{2+}$), which in turn lead to a complex strain field in the bulk of the material.

As to an additional complication, we refer to publications exploring intrinsic variation of iron concentration inside the (MgFe)O [13,14]. The experimental evidence suggests that iron cations tend to cluster forming inhomogeneous regions within crystal matrix. Obviously, the latter effect also correlates with spin state crossover and may be superimposed with other effects described above. All the mentioned observations indicate the complexity of spin state crossover and highlight (MgFe)O as an intricate and an underexplored system.

X-ray diffraction is the natural methodological choice for investigations of material strain and related effects, caused, for example, by subtle processes such as the spin state crossover. Considering different contributions to the diffraction signal and harsh experimental environment in high pressure diamond anvil cell (DAC) experiments, the sought information can only be provided from a single crystal of exceptional quality which should remain preserved during the compression. In comparison with powder material, single crystal diffraction signal of (Mg,Fe)O should be free from additional undesirable micro strains unavoidably appearing in compressed powders (grain boundaries, grain-grain contacts, higher concentration of dislocations, etc.). Importantly, for any existing diffraction condition, single crystals provide the highest signal-to-noise ratio if we consider the same sample scattering volume.

In this paper we investigate effect of high pressure on (Mg,Fe)O single crystal $(111)_{hkl}$ Bragg reflection by means of synchrotron based X-ray diffraction. We use the advantage of high resolution (controlled in our case by a combination of a long sample to detector distance (SDD) and a choice of X-ray energy) to monitor the effects of compression on sample diffraction signal. We analyze the signal collected before, during, and after spin state crossover. We pay special attention to a characterization of disordered state (spin state crossover) and, the material's response originating from the effects on the nano-scale.

Our study could also be considered important for other methodological reasons, particularly for developments in the field of dynamic compression experiments using diamond anvil cells. The findings presented below may provide references for analysis and comparison of the effects of the stresses and strains (and additional factors like relaxation processes) in relatively slow static and dynamic DACs experiments (membrane and piezo-driven mechanics).

## II. EXPERIMENT DETAILS

(MgFe)O single crystals were synthesized at Bayerisches Geoinstitut, Bayreuth University. We used a crystal from the batch already reported before [13], and we refer to the iron concentration of $x_{Fe}=0.19(1)$.

The small crystal (10 x 15 x 5 µm$^3$) was preselected at the P02.2 extreme conditions beamline [15], Petra-III, DESY using the energy of 42.7keV. Sample was loaded into a symmetric diamond anvil cell (DAC) with opening of 70° equipped with Boehler-Almax diamonds. We used Re as gasket material, Ne as pressure medium; small sphere of ruby (pressure marker) and tiny piece of tungsten were placed inside the sample chamber. Tungsten was used for the sample alignment to the X-ray beam and for the centering of the DAC to the goniometer rotation axis (strong X-ray absorber). The diamond anvil cell (X ray aperture ±32°) was manually pre-oriented in such a way that (MgFe)O $(111)_{hkl}$ was located in the horizontal plane.

X-ray diffraction data at high resolution were collected at the P10 Coherence Applications Beamline, Petra-III, DESY (Fig. 1). The energy of 15keV was selected by a channel cut monochromator. We used a stack of compound refractive lenses located 1.6m upstream from the sample in order to focus X-ray beam down to 2.5×4.5 µm$^2$ (vert. × hor.). The sample to detector distance was 5m with a long flight-path under vacuum installed in the horizontal plane between the sample and the Eiger X 4M detector (pixel size 75×75 µm$^2$). Bragg peak intensity was collected by rotating the sample along ϕ axis of a goniometer (scanning of a Bragg spot in the horizontal plane). At 15keV we were efficiently measuring the sample absorption and used this signal for precision centering. The calibration of *2θ* range at different detector position was done by performing a cross correlation with the position of Au powder (particle size <100nm) measured separately at ambient conditions at different detector positions. For this, we used a calibration by the position of Au $(111)_{hkl}$ collected at ambient conditions. Afterwards, the *2θ* angle of the flight-path tube was fixed, we adjusted the position of the detector to enhance the covering of (MgFe)O $(111)_{hkl}$ Bragg spot by the detector. This was necessary due to the presence of intrinsic gaps of sensitive material in the detector.

The choice of the Bragg reflection comes from following consideration: the $(111)_{hkl}$ is the plane with the densest packing, and we could expect the strongest manifestation of the crossover; at the same time this reflection is accessible through the aperture of DAC for the given, rather low, energy of 15 keV.

We have recorded the pressure dependence of the Bragg spot by scanning the ϕ axis (Fig. 1) and capturing the Bragg spot in slices of 0.01° step. Here and below we will present and refer to the data as "integrated" (or "integral intensity" as a full sum of individual slices resulting in the integral intensity of the Bragg spot, integrated over a range of ϕ) and as "slices" (data collected at a specific ϕ position, representing an individual slice of the Bragg peak). An example comparing the difference between slices and the integral intensity at 72 GPa is provided in Fig. 2. In that figure we also indicate the convention used for the analysis of the data as detected in the detector plane.

The Fig. 2 deserve additional clarification due to the presence of specific features appearing on the detector for the partially coherent incident beam. The high intensity speckles shown in Fig. 2 indicate a strong inhomogeneity of scattered intensity detectible in both integrated intensity and slices. The portion of the coherent fraction was significant enough that we can observe additional features in form of interference fringes. However, the coherent fraction used and the oversampling conditions do not allow to use phasing algorithms to fully reconstruct the sample. The integrated scattering signal is unaffected by the conditions of partial coherency. Thus, the coherent beam fraction was not sufficient to affect the main scattering signal of the measured Bragg spot.

We suggest that some of the new features could be attributed to the high pressure environment surrounding the sample. Indeed, the phase of the partially coherent X-ray beam can modified during the traversing through the sample environment. The properties of X-ray beam can be modified prior to the diffraction of the beam by (MgFe)O, during and/or after. The important fact is that certain features related to the X-ray beam path on the incident and the scattered intervals can manifest to the final Bragg spot in terms of fringes of weak intensity. For example, we can introduce strong contrast between different materials, shape of the sample, grain boundaries of material with strong scattering power and etc. Since the fringes signal is weak, they do not interfere with our analysis of the Bragg spot which we perform within the framework of conventional diffraction.

In the case of the current study, we see a nice correspondence of the Ne grain boundaries (visible light) with additional lines appearing on the Bragg spot (Fig. 2, see also Fig. B1 of Appendix B). The contribution of Ne grains to the diffraction signal of (MgFe)O is highly probable. We note, that Ne pressure medium has significant volume and strong scattering power. We also remind, that Ne contribution to the (MgFe)O signal cannot be considered a diffraction signal from the Ne itself, and in order to finish this discussion, we report that after crystallization of Ne, its major grain boundaries remained at the same positions and changed only as far as could be expected from a conventional diamond anvil cell experiment.

Before we present the results, we would like to have a brief introduction of the methodology of the data collection. By scanning the Bragg spot by means of the 2D detector and the $\phi$ motorized rotation stage (see also Fig. 1 (Right) for clarification), we selected to trace the following practical parameters: center of mass and maximum intensity positions of the Bragg spot peak ($2\theta$ angle, or d-spacing), width of the spot profile as projected on the arc of diffraction cone ($\chi$ angle) and width of the spot profile as a function of $\phi$ rotation. Data on d-spacing and width of the peak in the detector plane perpendicular to the diffraction cone (along $2\theta$), represents data on the current average stress conditions and strain. Data on the $\chi$ angle should produce information on the mosaicity of the crystal, if it would be evolving under pressure, but it should not be considered limited to it.

The size of the particular Bragg spot in the reciprocal and the direct space is affected by many parameters, including the instrument. In our case it is a natural choice to use the width of the Bragg spot in units of $\chi$ and $\phi$ angle rotation to detect a trend of Bragg spot broadening due to an additional scattering which could be attributed to the presence of spin state disorder in the material caused by the crossover. As would be clear from the results, a single d-spacing or $2\theta$ projection is not fully suitable for this purpose.

### III. RESULTS

We collected X-ray diffraction data at several pressure points, covering the pressure region before, during and after the spin state crossover. The resulting pressure-volume (P-V) relation is plotted at Fig. 3 together with some reference equations of state (EOS) for pure MgO.

In the insert of Fig. 3 we show $2\theta$ profile width of the integrated $(111)_{hkl}$ Bragg spot and illustrate our definition of the Bragg peak Full Width ($FW_{1/3}$), which we calculate here at $1/3^{rd}$ of the peak height ($2\theta$ and $\chi$ projections). Special definition of these parameters is necessary because the shape of the peak was asymmetric at all pressure points (effect of high-resolution X-ray diffraction). Due to the specific shape of the peak, for each pressure point we present two values of the lattice parameter and the unit cell volume - one for the center of the peak mass (open symbols on the figure) and the position corresponding to the maximum intensity (solid symbols).

The unusually broad intensity profile of the peak in $2\theta$ projection requires a separate clarification. It is true for most of the common 2D large area detectors used at popular high pressure diffraction beamlines, that their pixel size ranges from 144um to 200um (detectors with small sensitive area like Mar165 or

Lambda are not considered). In a case of SDD equal to 400mm and constant wavelength, the projected width of the Bragg spot (integrated) would be at maximum 3 pixels for the highest width of the Bragg spot reported in this paper (at highest pressures and conventional X-ray diffraction energy of 32-42keV). Thus, by the criteria commonly used in the high pressure community, the data reported here was collected from crystal of (MgFe)O of exceptional quality, which was preserved at all experimental points, as it is possible in a diamond anvil cell high pressure environment.

Going back to Fig. 3, we see that our data is in good agreement with the literature [16–18]. Considering previously reported equations of state for the pure MgO material, we note some inconsistencies among them with the strongest one related to nanosize MgO material [19]. Here and below we will use the data of Jacobsen et al. [18] measured on a small single crystal of MgO in He pressure medium at room temperature, which we consider as a reference EOS for bulk material.

In Fig. 4 we compare our P-V data with the equation of states for MgO by calculating a difference between unit cell volume of our material and the unit cell volume of MgO at the same pressures [18]. The region of spin state crossover is clearly visible and agrees well with previous measurements [13,20,21].

Having confirmed the region of the spin state crossover, we present the data describing (MgFe)O $(111)_{hkl}$ Bragg peak shape using three cross-sections, namely, using profiles collected in $2\theta$, $\chi$ projections and a profile collected during $\phi$ rotation as was introduced previously. Before we continue our analysis, for a complete description of the data shown in Fig. 3 and Fig. 4, we will discuss the point of 51 GPa, which was measured twice.

The reason for emphasizing this specific point comes from its vicinity to HS-LS crossover onset [13] for the specific composition of $x_{Fe}$=0.19(1). The first measurement corresponds to the time interval of 30-40min after compression to the target value, and the next data-set was collected 9 hours later ("relaxed" - pink symbols in figures). As one can already see from the Fig. 5 featuring $2\theta$ cross-section, the peak width of "relaxed" point is much narrower if compared with the other points and indicates stress and strain "relaxation" in the pressure chamber. These observations are also supported by the raw data displayed in Fig. B1 of Appendix B.

In Fig. 5 we can see that apart from the obvious compression effect influencing the Bragg peak position and intensity distribution as a function $2\theta$ (strongly asymmetric peak profile), it is really hard to spot a specific trend. Still, the relaxation effect at 51 GPa is quite obvious. One can clearly see that the width of the Bragg peak expressed in $2\theta$, or in convenient `strain` d-spacing units, became narrower indicating strain relaxation. The center of mass for this peak moved to the higher d-spacing by an insignificant value as illustrated in the same figure. At the same time, in contrast to the center of mass, the maximum of the intensity moved to the lower d-spacing. Since the positions for the center of mass and the maximum of intensity should have a strict relation with pressure dictated by the material equation of state (P-V relationship) we suggest that the strain "relaxation" is accompanied with stress "relaxation" effects in the sample chamber. Since the pressure measured using the ruby marker (external to the sample itself) has not changed within the errorbars, we consider this observation as a typical manifestation of complex stress-strain conditions of high pressure environment effects, typically disregarded in many of the experiments, but at the same time important. Our observations clearly indicate that a physical system compressed in a diamond anvil cell may require significant amount of time to reach equilibrium, especially at conditions of suppressed kinetics. It may well be a longer period of time per point than is typically granted for diffraction beamtimes at large scale synchrotron facilities.

Observations made for our particular sample may be applicable for any other sample compressed using diamond anvil cells. Considering the width of the measured profiles in d-spacing units (Fig. 5), we observed significant strain evolution during the time interval following the mechanical compression and equilibration. Although prior and after the 9 hours of relaxation, the pressure was remaining almost the same (position of the Bragg peak center of mass and frequency of ruby fluorescence signal), we emphasize that the crystal had strain distribution, and this distribution was much higher in the state preceding the relaxation. Although in our experience the exact values will strongly depend on sample size, and other conditions related to compression, in our particular case we report a change of $2\theta$ projection width in d-spacing units ($\Delta d/d$ at $FW_{1/3}$) from ~$4.5(4)^{-4}$ to ~$0.8(2)^{-4}$ for the cases preceding and following the relaxation, respectively. Here we are assume that relaxation processes occurs in the sample being an integral part of sample environment, and we assume that the soft pressure medium such as Ne, would have shorter relaxation times than the oxide system.

Our observations indicate that ambient temperature dynamic compression experiments, which become more and more popular (i.e. Ref. [22]), may not reach a full equilibrium at the time of the compression, and their comparison with the static experiments should be done carefully. Significant amount of strain accommodated during dynamic compression may divert the process from equilibrium path. If one considers the example of (MgFe)O, any inhomogeneous strain on powders (i.e. directional) may promote local distortions of initially perfect $FeO_6$ octahedron (tetrahedral or rhombohedral distortions) and either trigger the transformation of HS into LS prematurely (Ref. [21]) or delay it. Indeed, the criterions for HS to LS transition for ferric iron would be different depending on the distortion of local environment and the exact crystal field configuration. Local strains at dynamic compression will regulate boundaries of the spin state crossover stability field and it is almost impossible to control or characterize them.

If we consider the projection along $\chi$ direction, the broadening of the Bragg peak (Fig. 5) increases as we enter the region of the spin state crossover at 51 GPa. The corresponding width value reaches maximum at about 59 GPa and then decreases, approaching the typical values preceding the HS-LS crossover at 72GPa.

It is inherently difficult to present and describe evolution of the 3D Bragg peak as a function of pressure, especially in the situation when the Bragg peak shows additional interference effects (Fig. 2). On the right side of the Fig. 6 we demonstrate the evolution of the $(111)_{hkl}$ width in the direction perpendicular to the detector plane. The peak profile measurement during the $\phi$ rotation could be introduced in the framework of $\omega$-$2\theta$ scanning on conventional X-ray laboratory diffractometers equipped with point detector. Here, in order to avoid any confusion and in order to have an optimal representation, we use a different notation for the $FW_{15\% \phi}$ of the peak, namely, the width measured at 15% of the height.

In the Fig. 6, at the right panel, we see that the profile width increases as we approach the pressures of 51 GPa, and it then remains large up to 72GPa, namely the pressure range corresponding to the end of the crossover. The left panel of the Fig. 6 shows the $2\theta$ position of maxima for a given detector frame changing as a function of ϕ rotation. This representation reflect corresponding strain distribution in direct and reciprocal space. We remind the details of the scattering geometry by referencing Fig. 1 and the complementary illustration shown in Fig. A1 of Appendix A.

We analyze different projections and their widths in the following section, where we compare all our observations and discuss effects potentially responsible for the observed broadening of a Bragg spot. But, before doing that, we would like to provide additional practical information and we present data in Appendix C illustrating stability of the signal during the data collection. As can be seen from the figures shown in Appendix C, our signal is only weakly dependent on the top-up mode on the synchrotron. We investigated temporal stability of the Bragg spot central slice. We note, that the integral intensity of a Bragg spot is constant as well as the corresponding pattern recorded by the detector. The width of the Bragg spot does not change over the time of 80 seconds of data collection. This test indicates that the timescales of stress-strain relaxation effects described above are significantly longer than the 80 seconds of the test conducted.

## IV. DISCUSSION

Analysis of unit cell volume of (Mg,Fe)O as function of pressure (Fig. 4) suggests that spin crossover started at 51 GPa, in a good agreement with Ref. [13]. We compare the observations extracted from the Bragg peak projections in Fig. 7, and we report that $(111)_{hkl}$ Bragg reflection is sensitive to the spin state crossover from both: the position and size/shape point of view.

While effects of spin state crossover on $FM_{1/3\ d\text{-space}}$ ($2\theta$ projection) could be potentially influenced by the effects of 'stress-strain' relaxation described above, variation of $FW_{1/3}\chi$ ($\chi$ projection) and $FW_{15\%\ \phi}$ ($\phi$ projection) as a function of pressure indicate an obvious response to the crossover. The spin state crossover started at 51 GPa. Our observations are supported by the increase of $FW_{1/3}\chi$, especially pronounced after 9 hours relaxation at 51 GPa, and significant increase of $FW_{15\%\ \phi}$ at the same pressure. At pressure of 72GPa, at the end of the spin state crossover stability field, the value of $FW_{1/3}\chi$ decreased to the values preceding the crossover. Please also note that the values of $FW_{15\%\phi}$ saturate at pressure which should correspond to the middle of the spin crossover region.

The observed behavior of $FW_{1/3\chi}$ can be attributed to additional scattering arising from the appearance of the disorder and competition of HS/LS states. The observation that the maximum of $FW_{1/3\chi}$ corresponds to the middle of the crossover stability field supports our argument. This pressure region corresponds to the case of 1:1 abundance of HS and LS domains. If the changes of $FW_{1/3\chi}$ would be related with any enhancement of mosaicity in $\chi$ direction, the $FW_{1/3\chi}$ value would not decrease at higher pressures and exhibit the behavior close to that observed for $FW_{15\%\ \phi}$.

In turn, the changes of $FW_{15\%\ \phi}$ could be attributed to the defect-structure modification of the material. Within the restriction of fixed sample composition, we may consider formation of stacking faults and/or increased mosaicity of the sample in a specific direction. The scattering geometry (Fig. A1 of Appendix A) and the data shown in Fig. 6 suggest that the broadening appears in the direction close to be perpendicular to the detector plane. We can add that this is a direction perpendicular to $\chi$ projection, and the broadening or shape change occurs parallel or slightly inclined to the vector [111] in the reciprocal space, in such way that the Q scattering vector is constrained.

First of all, to our knowledge it is the first observation of correlation between microstructure changes in the material and the spin state crossover observed at high pressures on a single crystal. Considering different type of defects, we can exclude formation of uniformly distributed vacancies, in the latter case we would not see specific directional behavior – the broadening of the Bragg peak would appear in all directions, and mostly at the baseline of the Bragg peak. It should also be weak [23].

Taking other effects into account, the formation of stacking faults is one of the most probable explanations for the observed behavior. Stacking faults are considered to be manifestation of one dimensional disorder. (MgFe)O is material with *fcc* lattice, and thus, may have stacking faults of deformation or twin type [24]. While the former type of defects lead to diffraction line broadening, the latter are responsible for effects of broadening and line shift. Additional information and mathematical description can be found elsewhere [24].

Another explanation for the observed effect would be a formation of sub-grain boundaries leading to a minor rotation of $(111)_{hkl}$ planes along the $\phi$ axis (see scattering diagrams featured at Fig. 1 and A1 of

Appendix A). One can imagine crystal bending as an example of such situation. If we consider that HS and LS domains homogeneously coexist though the bulk of the material, we mark this scenario to be a less probable. There is no obvious scenario relevant for the spin state crossover explaining why mosaicity would evolve in a certain direction and would not evolve at the same time in χ cross-section. We mention the bulk of the material, since X-ray diffraction was collected in Laue geometry, and thus, the scattered signal carries information about the full thickness of the sample. In contrast, stacking faults can be formed locally, and this process seems to be a more logical strain release mechanism for the contact strains appearing due to HS-LS domain interaction.

Formation of the stacking faults as planar defects could be correlated with information on the slip planes typical for cubic systems. The main slip systems for our material are $\{110\}(1\bar{1}0)$, $\{100\}(011)$, and $\{111\}(1\bar{1}0)$. Although, the slip systems of $\{100\}(011)$ and $\{111\}(1\bar{1}0)$ were shown to be less probable than $\{110\}(1\bar{1}0)$, as follows from a study employing strongly non hydrostatic conditions [4], we note that the latter study determined the most probable and dominant slip planes, and indeed, the less probable $\{111\}(1\bar{1}0)$ slip plane could still be relevant to the behavior attributed to the spin state crossover.

In order to avoid any confusion, we would like to clarify, that the observed broadening attributed to stacking faults would be observed differently for powder and single crystalline material. In powders, the response is averaged over an ensemble of many grains and numerous orientations, while in a case of a single crystal the effect should be directional. Due to the specific limitations of our experiment we could access only one Bragg peak and could not retrieve additional information. Finally, one should not consider the picture of defects, or stacking faults formation as the only mechanism accommodating the contact strains of HS-LS during the crossover. Still, to our knowledge this is the first experimental observation of such process. We can mention other examples related to the HS-LS contact stain accommodation which have not yet been given attention from the high pressure community. They include HS-LS boundary modification of $Fe^{2+}$ local crystal structure changing local crystal field configuration, and thus, electron distribution across different $d$-electron orbitals. Such effects are very difficult to resolve experimentally in conventional high pressure X-ray diffraction and also present a certain challenge to theoretical studies.

All in all, at this point we can conclude that data collected in the χ projection of (111)$_{hkl}$ Bragg peak demonstrates additional finite scattering correlated with the disorder caused by presence of HS and LS. We also report formation of static structural defects which accumulate as material is transformed domain by domain from HS to LS.

We consider that the discussion would not be complete if we do not address temporal evolution of our signal for the individual slices of the Bragg peak. In the Introduction we mentioned the consideration, that in a situation of spin state crossover and strongly suppressed kinetics, the system may still `breath` by switching HS to LS – the process of small to little energy fluctuation (e.g. temperature). This process could be potentially investigated if we inspect and compare evolution of diffraction spot speckles collected before, during and after the spin state crossover as a function of time. Data presented in Appendix C allows us to conclude that within 80 seconds of data collection with a frame exposure time of 0.05 second, we do not resolve any additional features which could be attributed to the HS LS domain fluctuations. If we will slice the Bragg spot at the center and cross correlate variations of different regions of interest with finite intensity, then all the anomalies detected in the temporally resolved signal will be unambiguously attributed to the fluctuations related to t.he operation of the synchrotron source. This observations indicates that the process of domain 'breathing' is either much faster than the exposure of the frames, and here we limited by the photon flux, or, which is equally probable, the process is much slower and could be resolved in future studies.

## V. CONCLUSION

By using (MgFe)O as an example of a material with strongly correlated properties and with the distinct, relatively broad spin state crossover of ferrous iron we characterize the latter phenomenon employing high resolution X-ray diffraction under high pressure conditions.

We select the composition with moderate iron content $x_{Fe}$=0.19(1) and focus on the features of (111)$_{hkl}$ Bragg spot as we approach and pass through the pressure range of coexisting HS/LS states. We find and describe the effects of stress-strain relaxation as well as subtle changes of Bragg spot shape and size, some of those we attribute to the formation of static microstructure defects accumulating during the spin state crossover in a single crystal material. We suggest that the formation of microstructural defects is one of intrinsic processes of contact strain relaxation upon transformation of HS domains to LS.

Finally, even for a composition of $x_{Fe}=0.19(1)$, we detect an evidence which we describe as the additional contribution to the diffraction signal caused by the presence of disorder in form of HS-LS domains. Indeed, this contribution, manifested through an additional Bragg peak broadening reaches maximum at pressures corresponding to the middle of the crossover, and thus, it corresponds to the equal abundance of the HS and LS.

## ACKNOWLEDGMENTS

We acknowledge DESY (Hamburg, Germany), a member of the Helmholtz Association HGF, for the provision of experimental facilities. Parts of this research were carried out at Petra-III, DESY. We acknowledge Extreme Conditions Science Infrastructure of Petra-III, DESY for their help and access to their visible light spectroscopic facilities.

# FIGURES

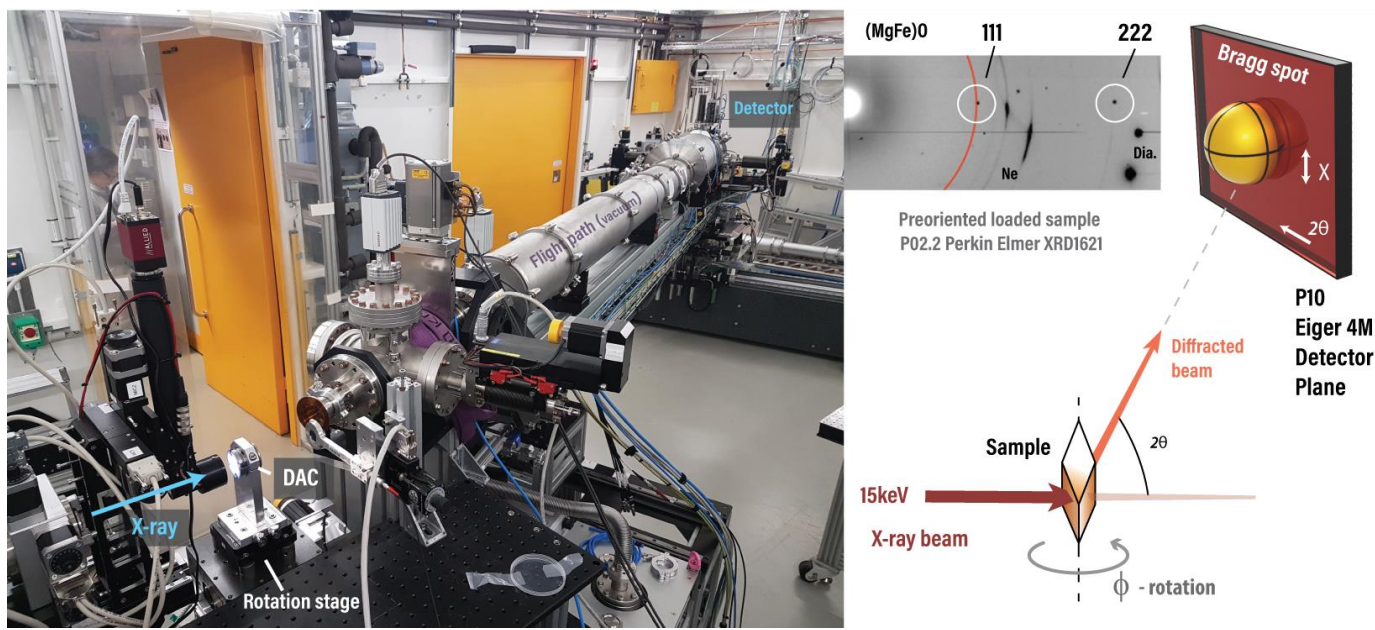

FIG. 1. (**Left**) A photograph of the high resolution experimental setup of P10 beamline with a diamond anvil cell. (**Right**) A diagram supplementing the photograph indicating the geometry of collection used in the study. At a high resolution regime of the beamline we observe 3D shape of the Bragg peak as a function of 2θ, χ directions and as a function of ϕ rotation. Using a 2D pattern shown in the middle, we highlight quality of the loaded and pre-aligned sample. The latter pattern was collected at the beamline P02.2 as a preparation step. Data was collected at a different energy in comparison to P10. We mark the Bragg reflections $(111)_{hkl}$ and $(222)_{hkl}$ and indicate additional scattering from the sample chamber (Ne pressure medium and diamond Bragg peaks).

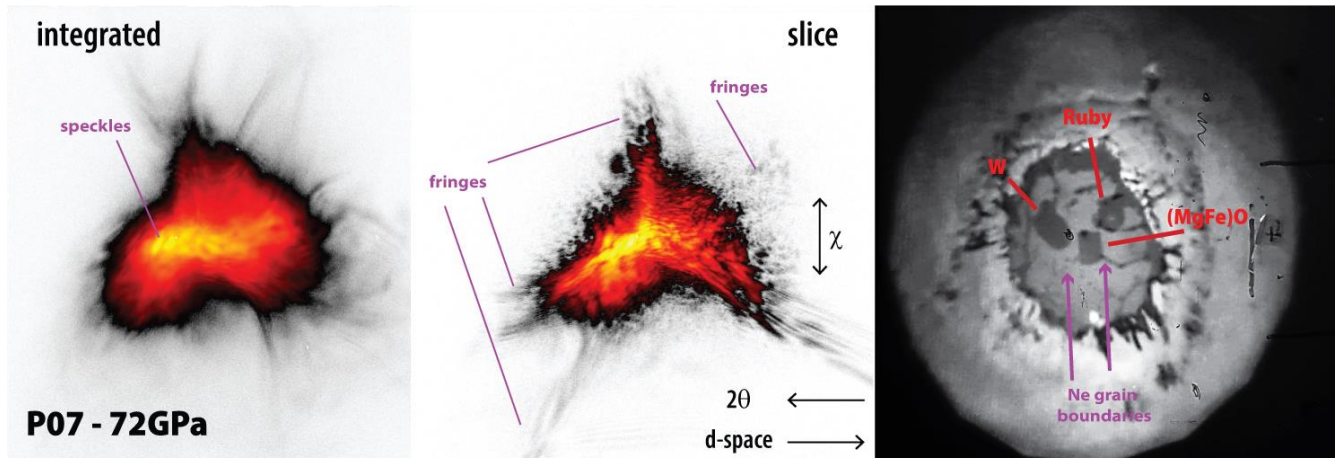

FIG. 2. Snapshots of the Bragg spot as captured by the detector and a photograph of the sample chamber at high pressure of 72GPa. (Left) Integrated Bragg spot featuring full intensity of the peak and additional speckle structure. (Middle) Slice of the Bragg peak in the vicinity of the maximum intensity. We show the directions for the axes corresponding to the $\chi$ and $2\theta$ angles. $\chi$ corresponds to the angle on the arc of the diffraction circle. We highlight the specific contributions to the Bragg spot including the speckles of high intensity, additional interference fringes of weak intensity due to the non-zero coherent fraction of the X-ray beam. (Right) The photograph of the sample chamber made using visible light (with indirect backlight illumination). Diamond has a culet size of $200\mu$m. Note the striking resemblance of the Ne grain boundaries with the features on the snapshot in the middle.

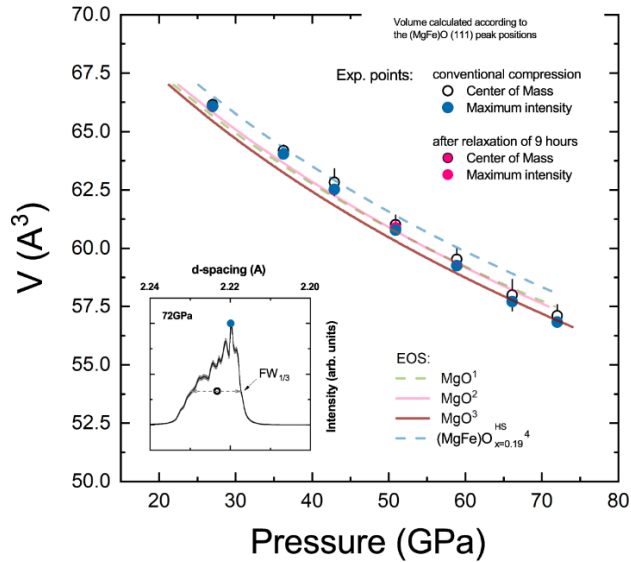

FIG. 3. Unit cell volume as a function of pressure for our (MgFe)O material. Solid points correspond to the maximum intensity of a Bragg peak, while open points correspond to the center of the Bragg peak mass. The insert demonstrates a profile of the full integrated Bragg spot intensity as a function of d-spacing. In order to capture the width of the peak by an optimal way, here, we define the Full Width at $1/3^{rd}$ of the peak height. For the guidelines we use equation of states (EOS): 1 - [16], 2 - [17], 3 - [18] , 4 - [13] with the latter indicating the HS EOS.

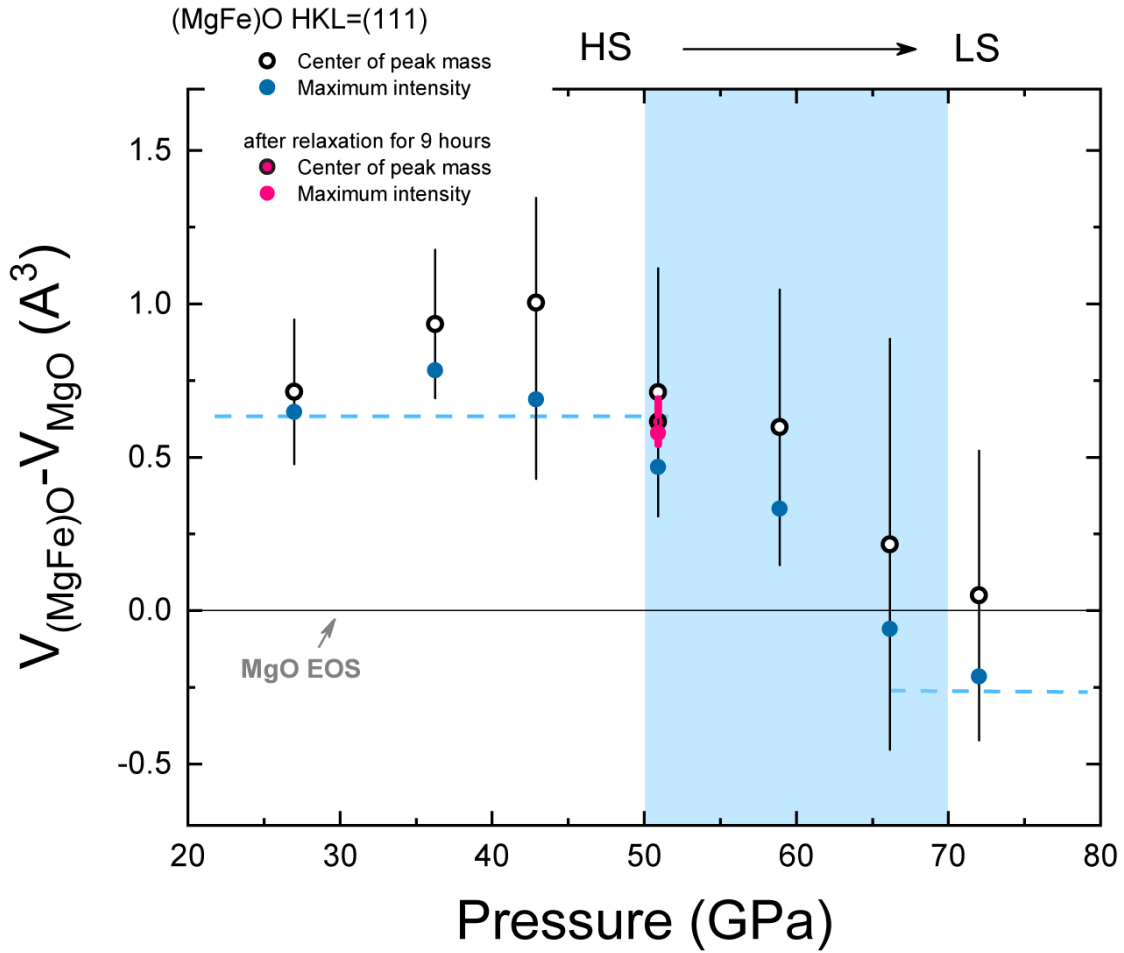

FIG. 4. Difference of (MgFe)O and MgO unit cell volumes as a function of pressure. Open symbol were calculated using the Bragg peak center of mass, while the filled symbols were calculated using the maximum of the Bragg spot peak. Error bars represent the half width at $1/3^{rd}$ of the Bragg peak width as introduced in Fig. 3. MgO equation of states is taken from [18]. Dash lines are the guidelines. The tentative region of HS to LS crossover is indicated with blue color.

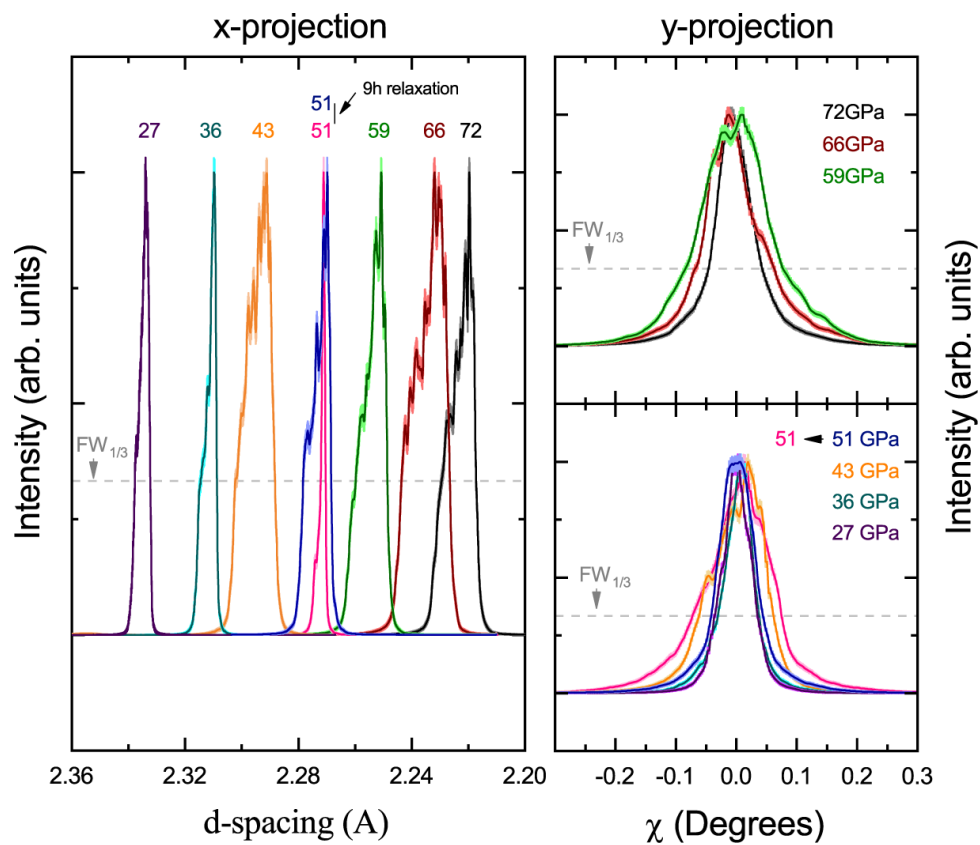

FIG. 5. Projections χ and 2θ of the Bragg spot on as a function of pressure. Intensities are normalized. 2θ is represented using the units of d-spacing. (Left) The projection of integrated Bragg peak on 2θ axis slides to the direction of lower d-spacing values as a result of compression. We highlight the point of 51 GPa. This point was measured twice with 9 hours between the measurements. After hours of relaxation, the width of the peak in 2θ has become narrower. Our observation indicates a complex stress-strain relaxation process. The 2θ position of the peak did not changed significantly while the corresponding projection on χ increased after relaxation period. (Right) The projection of integrated Bragg peak on χ. The width of the peak in χ projection has increased upon entering the stability region of the spin state crossover, but then it decreases as we compress further and complete the transition at 72GPa.

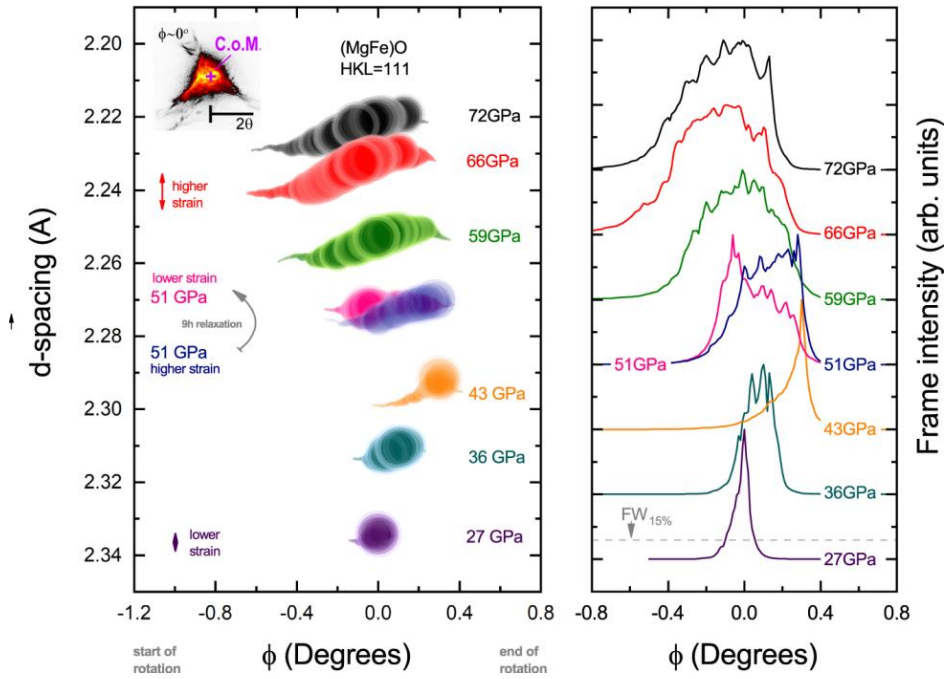

FIG. 6. (Right) Width of the Bragg spot in the direction perpendicular to the detector plane as a function of pressure. It is provided in the units of sample rotation along the ϕ axis. For simplicity or representation, we assign ϕ to 0° in the vicinity of the Bragg spot central slice. Full Width (FW$_{15\%}$ ϕ) is estimated at 15% of the normalized peak intensity. (Left) Intensity and $2\theta$ position of the Bragg spot as a function of pressure and ϕ rotation. Data represents changes of intensity center of mass (C.o.M.) and its position measured as a function of ϕ rotation or a Bragg spot slice (see the insert). Larger and smaller spots indicate higher and lower intensity of the slice, respectively. The size of the points varies in linear scale as a function of the in-frame intensity, and they are shown in ascending sorting. Please note that the Bragg spot is not a simple ellipsoid, but has a certain non-trivial 3D distribution of intensity in the reciprocal space. At a given Bragg spot slice, the C.o.M. of intensity changes its 2θ position as a function of ϕ rotation reflecting scattering from a single crystal. There is an obvious increase of the width and a change of Bragg spot shape within the spin state crossover region. At lower pressures we see stronger confinement of intensity, which is spread out at higher pressures and reaches maximum at 66 GPa.

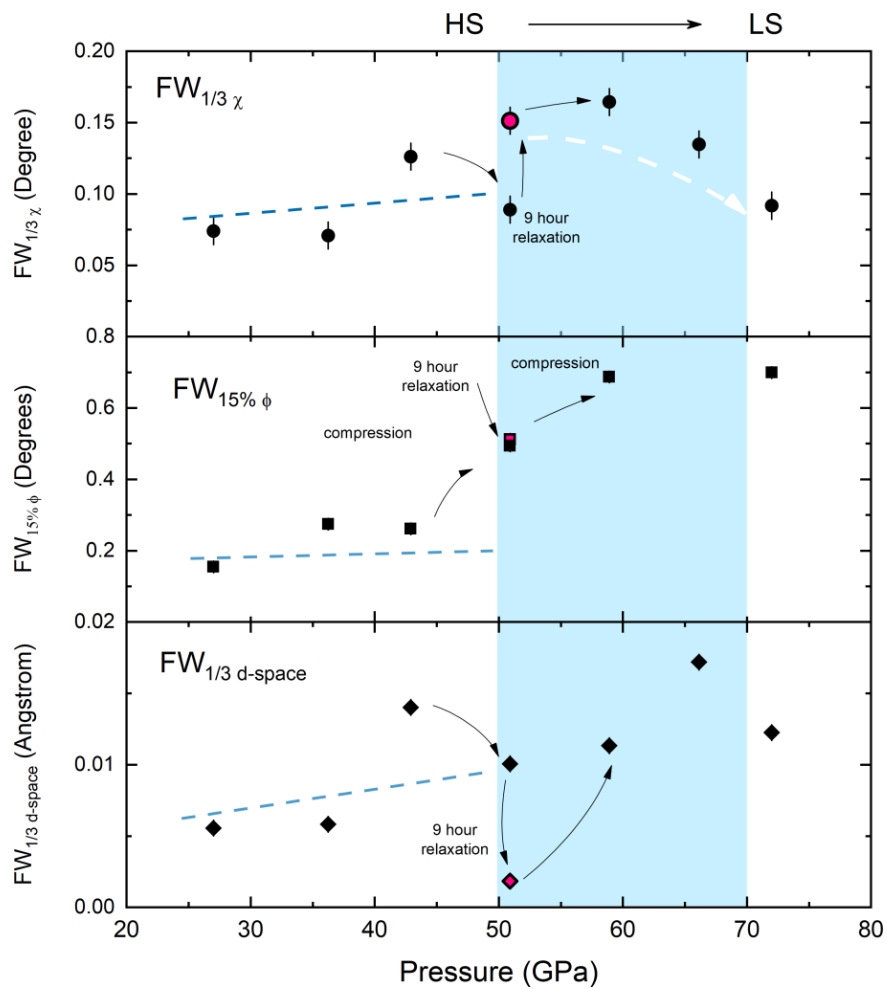

FIG. 7. Variation of Bragg peak $(111)_{hkl}$ Full Width (FW) of Bragg peak projection (Top) $\chi$, (Middle) $\phi$ and (Bottom) $2\theta$ under pressure. Note the increase of $FW_{15\%\phi}$ at the stability field of crossover and reduction of $FW_{1/3\chi}$ at pressures with sample almost fully transformed into LS.

# APPENDIX

## APPENDIX A: SCATTERING DIAGRAMS

In the Fig. A1 we demonstrate basic scattering diagrams supplementing our discussion on the geometry of the data collection and the Bragg spot broadening detected during $\phi$ rotation. The direction of the $\chi$ projection is perpendicular to the diagram plane.

If we would detect a broadening of the Bragg peak in the direction perpendicular to the detector plane, in the reciprocal space that would correspond to a change of shape in the [111] vector on a tangent to the iso-Q surface.

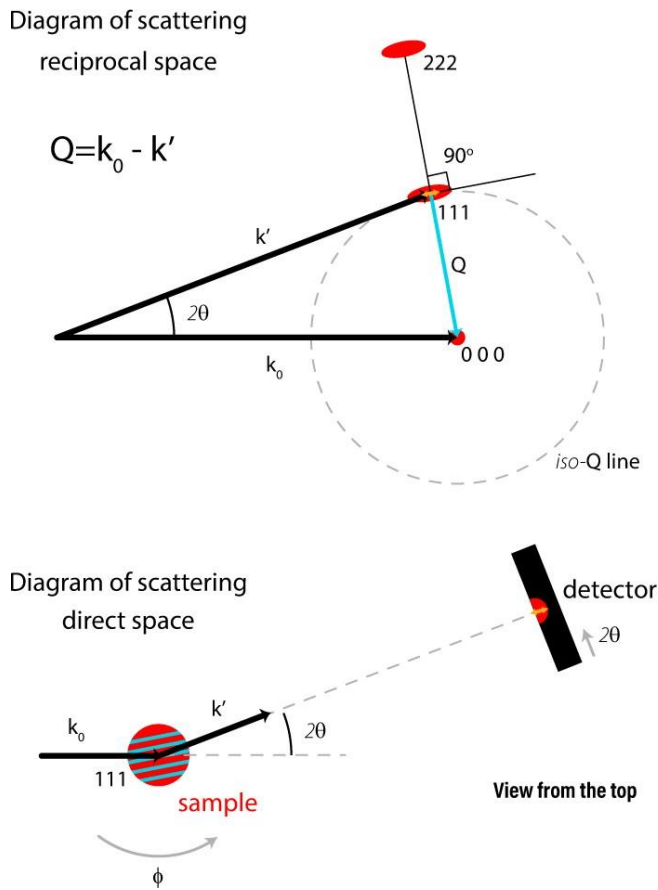

FIG. A1. Scattering diagrams describing experiment at the beamline P10. Please see text for discussion. Orange line indicates the situation of the of the Bragg spot broadening in the reciprocal space as detected during $\phi$ rotation in the direction perpendicular to the plane of the detector.

# APPENDIX B: INTEGRAL INTENSITIES OF (111)$_{HKL}$ BRAGG SPOT

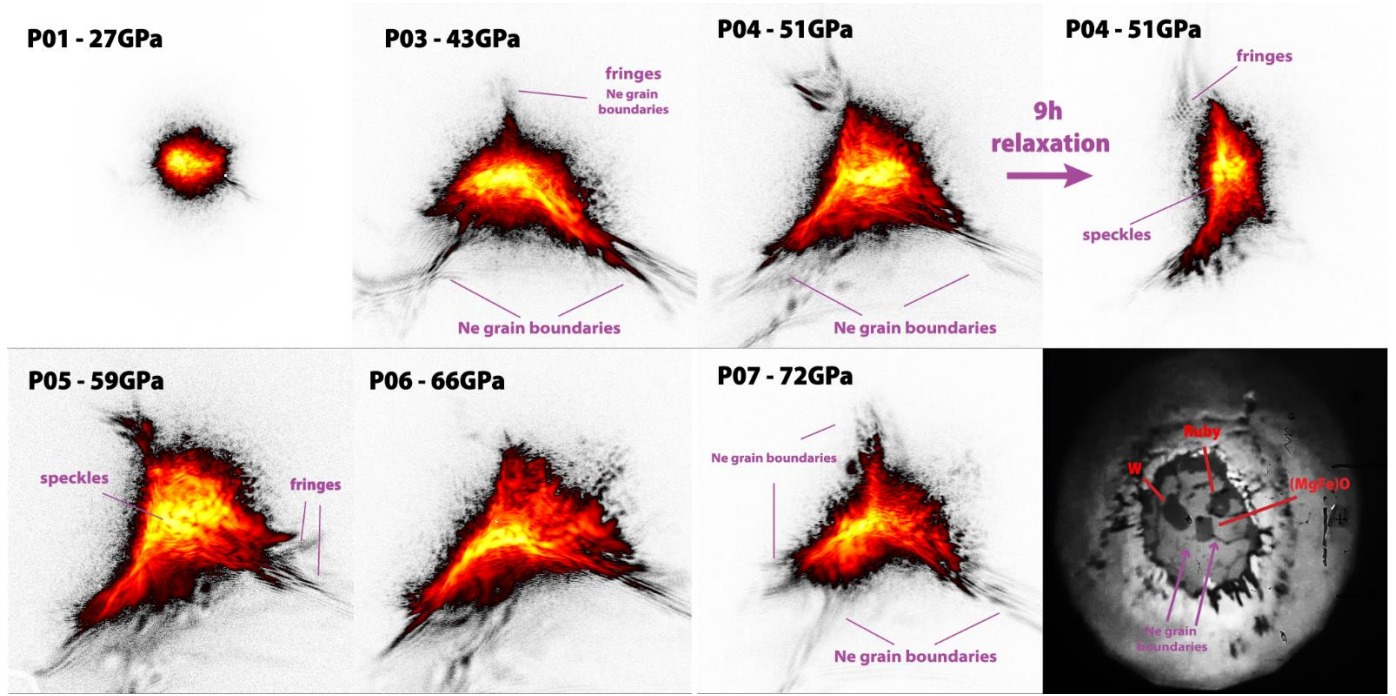

FIG. B1. Central slices of the (111)$_{hkl}$ Bragg spot collected for 80s together with an illustration of additional contribution to the diffraction signal.

# APPENDIX C: TEMPORAL PROPERTIES OF BRAGG SPOT

In the Fig. C1 we present our measurements of the frame total intensity stability as a function of time. It indicates that the section of the (MgFe)O $(111)_{hkl}$ Bragg spot was not considerably changing during 80s of collection time.

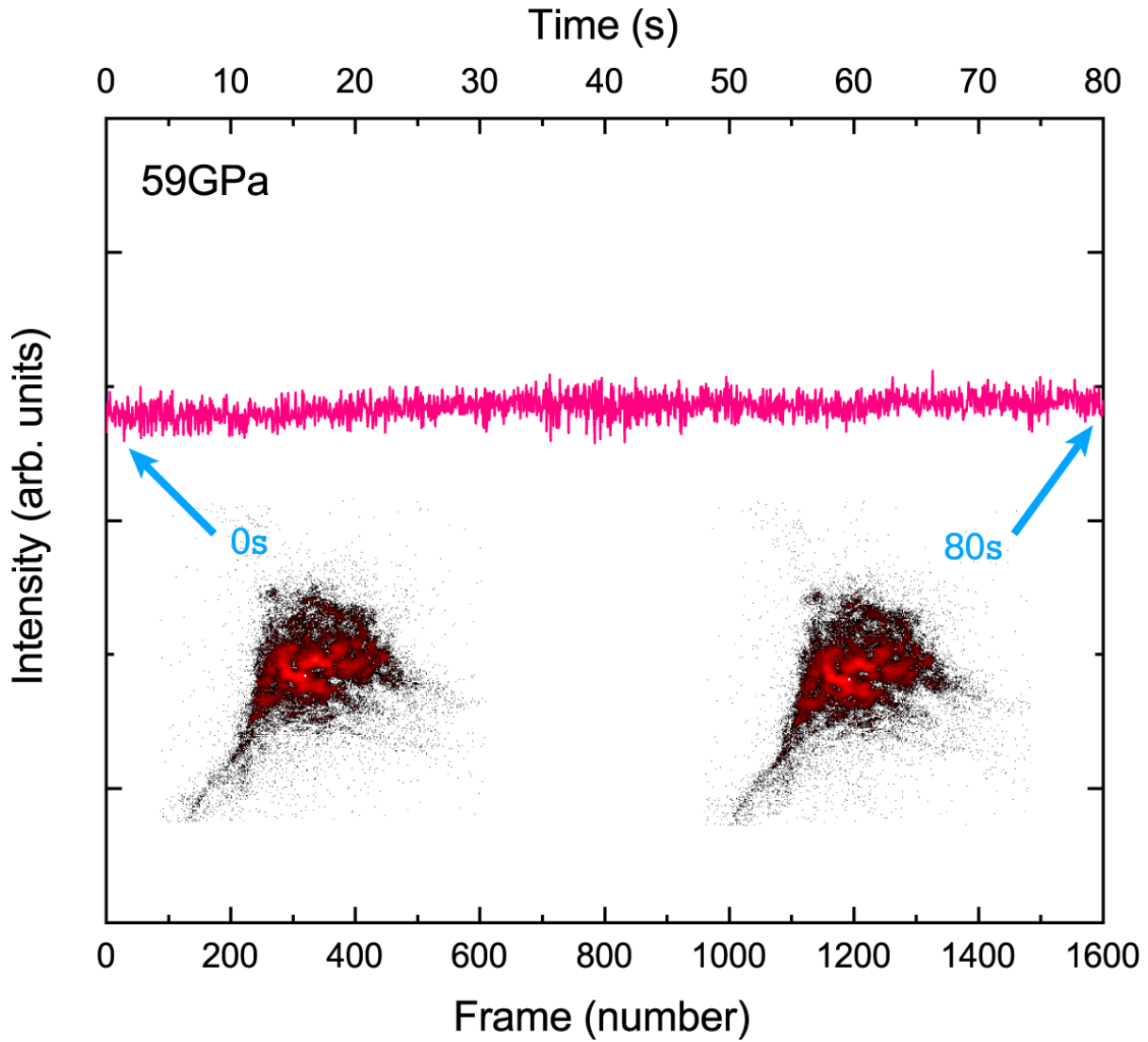

FIG. C1. Stability of the in-frame intensity as a function of time. Data was collected at a selected $\phi$ angle and represent intensity variation over time for a central slice of the Bragg spot. Within the counting statistics, the Bragg spot image seems the same and not affected by the stress-strain relaxation mentioned above.

In addition, we would like to mention importance of the synchrotron source stability on the measurements. This is indicated in the Fig. C2 where we compare intensity of selected speckles as a function of time in correlation with the Petra-III synchrotron current in 1% top-up mode. One can see small variation of intensity related with the synchrotron bunch refilling procedure, as we suggest related with small angular deviation of the incident beam falling on the sample. These small deviations did not affect our data on the widths of the Bragg spot.

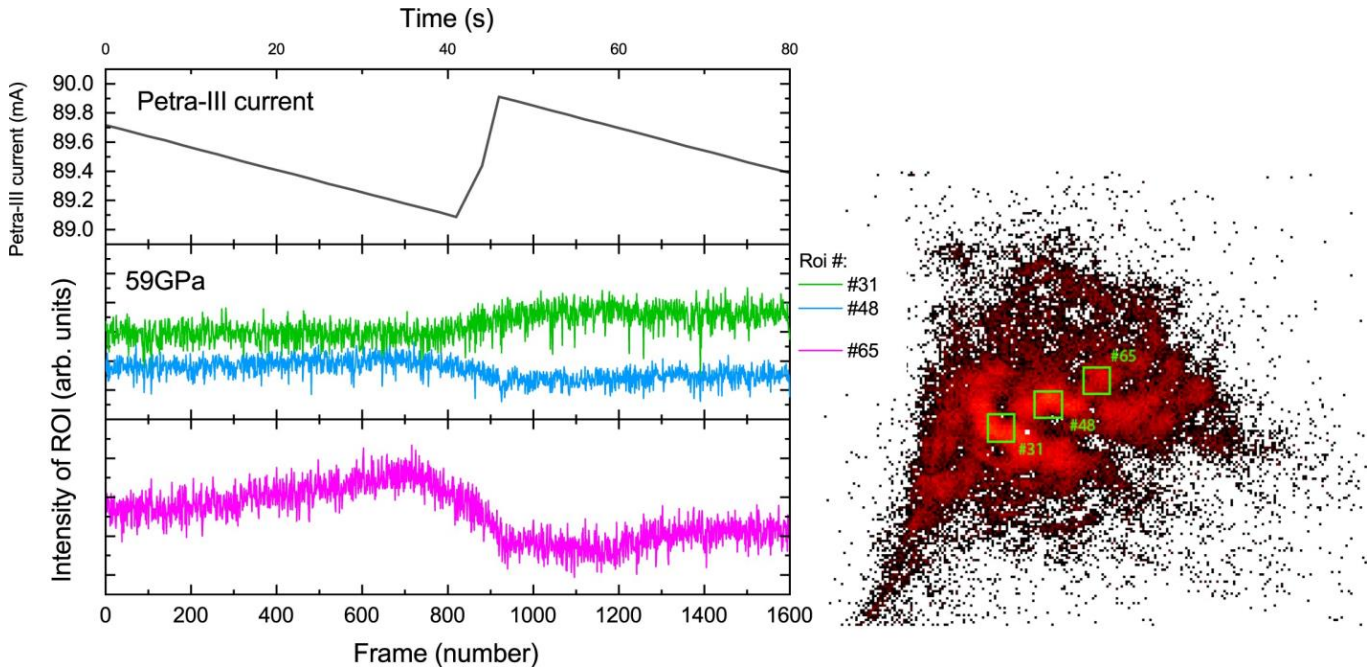

FIG. C2 Variation of selected speckle intensities (numbered regions of interest for the Bragg spot shown on the right) as a function of time. Data is shown in correlation with Petra-III current change. Data was collected at a specific $\phi$ angle and represent a central slice of the $(111)_{hkl}$ Bragg spot collected at pressures close to the middle of the HS-LS crossover.